\documentclass[copyright]{eptcs}
\providecommand{\event}{Workshop on Parallel and Distributed Methods in verifiCation PDMC 2009}
\providecommand{\volume}{??}
\providecommand{\anno}{2009}
\providecommand{\firstpage}{1}
\providecommand{\eid}{??}
\usepackage{breakurl}
\usepackage{graphicx}
\hypersetup{
  pdftitle={Tarmo: A Framework for Parallelized Bounded Model Checking},%
  pdfsubject={PDMC 2009 paper},%
  pdfauthor={Siert Wieringa, Matti Niemenmaa, and Keijo Heljanko},%
  pdfkeywords={Bounded Model Checking, Parallel algorithms, Distributed Algorithms}}

\renewcommand{\P}{\mathbf{P}}             
\newcommand{\Q}{\mathbf{Q}}               
\newcommand{\QP}[1]{\mathbf{Q'_{#1}}}     
\newcommand{\F}{\mathbf{F}}               

\newcommand{\FM}[1]{\F_{M\phi}^{#1}}      
\newcommand{\LD}[1]{\mathbf{LD}_{s_{#1}}} 
\newcommand{\LS}[1]{\mathbf{LS}_{s_{#1}}} 
\newcommand{\RWLock}{L}                   

\newcommand{\icnf}{iCNF}
\newcommand{\dip}{D}
\newcommand{\GQ}[1]{Q_{#1}^{\dip}}

\newcommand{\true}{\textbf{true}}
\newcommand{\false}{\textbf{false}}

\newcommand{\keyw}[1]{\textbf{#1}}
\newcommand{\FOR}{\keyw{for}}
\newcommand{\ENDFOR}{\keyw{end for}}

\newcommand{\LOCK}{\keyw{lock}}
\newcommand{\UNLOCK}{\keyw{unlock}}

\newcommand{\tarmo}{Tarmo}
\newcommand{\conv}{\textbf{CONV}}
\newcommand{\fourconv}{\textbf{4xCONV}}
\newcommand{\multiconv}{\textbf{MULTICONV}}
\newcommand{\multiconvsimple}{\textbf{MULTICONV-SIMPLE}}
\newcommand{\multiconvfullsharing}{\textbf{MULTICONV-FULL}}
\newcommand{\multiconvlesssharing}{\textbf{MULTICONV-ADAPTIVE}}
\newcommand{\multiconvtarmo}{\textbf{MULTICONV-TARMO}}
\newcommand{\multibound}{\textbf{MULTIBOUND}}
\newcommand{\multiboundtarmo}{\textbf{MULTIBOUND-TARMO}}
\newcommand{\multiconvbound}{\textbf{MULTICONVxMULTIBOUND}}
\newcommand{\distributed}{\textbf{DISTRIBUTED}}

\newtheorem{corollary}{Corollary}[section]
\newtheorem{example}{Example}[section]
\newtheorem{algorithm}{Algorithm}[section]

\newcounter{subfig}[figure]
\newenvironment{subfigcnt}[1]{\refstepcounter{subfig}\label{fig:#1}}{}

\newcommand{\fig}[3][width=\figwidth, angle=270]{\begin{center}\includegraphics[#1]{figures/#2}\caption{#3}\label{fig:#2}\end{center}}

\title{\emph{\tarmo}: A Framework for Parallelized Bounded Model Checking}
\author{Siert Wieringa \qquad\qquad Matti Niemenmaa \qquad\qquad Keijo Heljanko
  \email{Siert.Wieringa@tkk.fi \quad Matti.Niemenmaa@tkk.fi \quad Keijo.Heljanko@tkk.fi}
  \institute{Helsinki University of Technology TKK \\
    Department of Information and Computer Science \\
    P.O. Box 5400,~FI-02015 TKK,~FINLAND}
}
\def\titlerunning{\tarmo: A Framework for Parallelized BMC}
\def\authorrunning{S. Wieringa \& M. Niemenmaa \& K. Heljanko}
\begin{document}
\maketitle

\begin{abstract}
  This paper investigates approaches to parallelizing Bounded Model Checking (BMC) for shared memory environments as well as for clusters of workstations.
  We present a generic framework for parallelized BMC named \emph{\tarmo}.
  Our framework can be used with any incremental SAT encoding for BMC but for the results in this paper we use only the current state-of-the-art encoding for full PLTL \cite{DBLP:journals/lmcs/BiereHJLS06}.
  Using this encoding allows us to check both safety and liveness properties, contrary to an earlier work on distributing BMC that is limited to safety properties only.

  Despite our focus on BMC after it has been translated to SAT, existing distributed SAT solvers are not well suited for our application.
  This is because solving a BMC problem is not solving a set of independent SAT instances but rather involves solving multiple related SAT instances, encoded \emph{incrementally}, where the satisfiability of each instance corresponds to the existence of a counterexample of a specific length.
  Our framework includes a generic architecture for a shared clause database that allows easy clause sharing between SAT solver threads solving various such instances.

  We present extensive experimental results obtained with multiple variants of our \tarmo~implementation.
  Our shared memory variants have a significantly better performance than conventional single threaded approaches, which is a result that many users can benefit from as multi-core and multi-processor technology is widely available.
  Furthermore we demonstrate that our framework can be deployed in a typical cluster of workstations, where several multi-core machines are connected by a network.
\end{abstract}

\section{Introduction}
\emph{Bounded Model Checking} (BMC) is a symbolic model checking technique \cite{DBLP:conf/tacas/BiereCCZ99,DBLP:journals/lmcs/BiereHJLS06} which attempts to leverage the existence of efficient solvers for the \emph{propositional satisfiability problem} (SAT), so-called \emph{SAT solvers} (e.g.{\ }\cite{DBLP:conf/dac/MoskewiczMZZM01,DBLP:conf/sat/EenS03}).
SAT is the problem of finding a truth assignment to the Boolean variables of a propositional logic formula in such a way that the formula evaluates to \true, or determining that no such assignment exists.
This classifies the formula as respectively \emph{satisfiable} or \emph{unsatisfiable}.

The main idea behind BMC is to encode a system model $M$, property $\phi$ and integer $k$ called the \emph{bound} into a propositional logic formula in such a way that it is satisfiable iff there exists an execution of length $k$ of system $M$ which violates the property $\phi$.
Such an execution is called a \emph{counterexample}.
A conventional scheme for BMC is to have a SAT solver test the existence of a counterexample of length $k$, and if its existence is disproven (i.e.{\ }the solver returns ``unsatisfiable'') $k$ is increased after which the test is repeated.
A typical instance of this process is to start with $k=0$ and on every iteration increment $k$ by one.
The process ends whenever a counterexample is found or time or memory resources available run out.
We will call this approach \conv~for \emph{conventional}.
Notice that BMC in this basic form, to which we limit ourselves in this paper, is an incomplete method as it cannot prove a property $\phi$ correct for all possible executions of system $M$.
For a survey into complete BMC methods see Section 7 of \cite{DBLP:journals/lmcs/BiereHJLS06}.

Although SAT is an NP-complete problem current state-of-the-art SAT solvers can solve many instances of SAT efficiently.
Conventional SAT solvers are based on the DPLL framework \cite{DBLP:journals/cacm/DavisLL62}, which requires the input formula to be in \emph{conjunctive normal form} (CNF).
A propositional logic formula is in this form if it is a conjunction of \emph{clauses}.
A clause is a disjunction of \emph{literals}.
A literal is an atomic proposition, i.e.{\ }either a Boolean variable $x_i$ or its negation $\lnot x_i$.
Note that a clause is satisfied by a truth assignment in which any one of its literals is assigned the value \true, and a CNF formula is satisfied if all of its clauses are satisfied.
For the remainder of this paper whenever we speak of a formula we mean an instance of SAT in CNF.
Note that such a formula can be represented as a set of clauses.

A SAT solver based on the DPLL framework repeatedly selects an unassigned variable as the \emph{branching variable} which it assigns to either $\true$ or $\false$.
After this the solver searches for a satisfying assignment in the reduced search space.
If no such assignment exists the procedure backtracks and assigns the branching variable to the opposite value.

The default SAT solver used by \tarmo~is MiniSAT 2.0 without the simplifier \cite{DBLP:conf/sat/EenS03} but it can easily be replaced with any other \emph{conflict driven} SAT solver which supports \emph{incremental SAT}.
A conflict driven SAT solver derives, or \emph{learns}, new clauses as it is working its way through the problem's search space.
These \emph{learned clauses} can be seen as additional lemmas that help the solver to avoid parts of the search space that contain no solutions.
In a typical SAT solver the clauses of the input formula are kept in the \emph{problem clause database}, whereas the learned clauses are in the \emph{learned clause database}.

\subsection{Incremental SAT}
In a number of applications, including BMC, SAT solvers are used to solve a set of formulas that share a large number of clauses.
If we were to solve these independently each solving process may make the same inferences, expressed as learned clauses, about the common subset of the formulas.
To avoid this repeated effort it would be desirable to reuse learned clauses between the consecutively executed solving processes, which is what an \emph{incremental} SAT solver is good for.

\begin{example}\label{exampleSuperset} Assume that we wish to sequentially solve the formulas $\langle~ \F_1,~\F_2,~\dots,~\F_n ~\rangle$ for which $\F_i = \bigcup_{j=1}^{i} \P_j$, i.e.{\ }each formula $\F_i$ equals the union of the previous formula $\F_{i-1}$ and a new set of clauses $\P_i$.
  Exploiting the incrementality of the sequence to reuse learned clauses is easy in this case:
  We can simply place the clauses $\F_1$ in the solver, solve, report the result for $\F_1$, add $\P_2$ to the solver, solve, report the result for $\F_2$, add $\P_3$ and so on.
  All learned clauses remain logical consequences of the problem clauses throughout this sequence, so all learned clauses can be reused in consecutive runs.
\end{example}

Unfortunately, for most applications, including ours, it does not hold that each formula is a superset of the preceding formula as in Example \ref{exampleSuperset}.
If we want to solve two consecutive formulas we may not only need to add clauses to the solver, we also may need to remove some.
However, if we remove clauses from the problem clause database the clauses in the learned clause database may no longer be implied by the problem clauses.
The concept of \emph{assumptions} was first introduced in \cite{DBLP:journals/entcs/EenS03} and it offers a way around this problem.
Only a simple modification to a standard SAT solver is required; the addition of the possibility to solve the formula in the problem clause database under a set of \emph{assumptions}.
An assumption is simply a variable assignment.
We will show next why this is sufficient.

\begin{example}\label{exampleIncSAT} Assume again that we wish to sequentially solve the formulas $\langle~ \F_1,~\F_2,~\dots,~\F_n ~\rangle$ but now each $\F_i = \Q_i \cup \bigcup_{j=1}^{i} \P_j$, i.e.{\ }each formula $\F_i$ now contains a subset of clauses $\Q_i$ that is contained only in $\F_i$.
Let $\{~x_1,~x_2,~\dots,~x_n~\}$ be a set of \emph{free} variables, i.e.{\ }a set of variables that do not occur in any clause in any of the formulas in the sequence.
Let $\QP{i} = \{~ C_j \vee x_i ~|~ C_j \in \Q_i ~\}$.
Note that if $x_i$ is assigned the value $\false$ then formula $\QP{i}$ becomes equivalent to $\Q_i$.
If, however, $x_i$ is assigned the value $\true$, then formula $\QP{i}$ becomes equivalent to $\true$.
As $x_i$ occurs only in the clauses of $\QP{i}$ and its negation $\lnot x_i$ does not occur in any clause, the solver may freely choose to assign $x_i$ the value $\true$ unless we force it otherwise, which we may do by means of an assumption.

We proceed in almost the same way as in Example \ref{exampleSuperset}: simply place the clauses $\P_1$ and $\QP{1}$ in the solver, solve under the assumption $x_1 = \false$, report the result for $\F_1$, add $\P_2$ and $\QP{2}$ to the solver, solve under the assumption $x_2 = \false$, report the result for $\F_2$, add $\P_3$ and $\QP{3}$ and so on.

As we never actually remove a clause from the problem clause database, we do not affect the consistency of the learned clause database.
\end{example}

We use the BMC encoding of \cite{DBLP:conf/cav/HeljankoJL05,DBLP:journals/lmcs/BiereHJLS06} to generate the SAT instances.
For the remainder of this paper we will represent an encoded BMC instance as a sequence of formulas $\langle~ \FM{1},~\FM{2},~\dots,~\FM{n} ~\rangle$ for which $\FM{i} \subseteq \FM{k}$ for any $k > i$.
Furthermore there exists a corresponding sequence of variables $\langle~ x_1,~x_2,~\dots,~x_n ~\rangle$ such that $\FM{i} \wedge \lnot x_i$ is satisfiable iff there exists a counterexample of length $i$ against property $\phi$ in model $M$.

\begin{corollary}If~$\FM{i} \models C_j$~then for any $k > i$ it holds that~$\FM{k} \models C_j$.\label{cor1}\end{corollary}

From experiments in the early stages of this project we found out that it is not uncommon for the separate SAT instances in a formula sequence to take several minutes to solve while the whole sequence could have been solved using an incremental SAT solver in less than one minute.
The use of incremental SAT is thus crucial for performance when solving BMC instances, which makes general purpose distributed SAT solvers unsuitable for solving them.
In this paper we present approaches to parallelizing the solving of BMC instances while maintaining the efficiency of incremental SAT.
One of our main contributions is the introduction of a generic architecture for a shared clause database which allows sharing clauses between incremental SAT solver threads, allowing solvers to easily pick only those clauses from the database that are implied by their own problem clauses, while requiring only a small amount of bookkeeping. 
We demonstrate the feasibility of our design in environments where multiple solver threads can access shared memory, as well as for environments where solver threads communicate through a network.

In contrary to the approach presented for distributed bounded model checking of safety properties in \cite{DBLP:conf/fmics/AbrahamSBFH06} the correctness of our clause sharing mechanism is not dependent on the chosen encoding of BMC instances into incremental SAT.
Our framework can thus always benefit from future improvements in such encodings.
We chose to use the current state-of-the-art encoding presented in \cite{DBLP:journals/lmcs/BiereHJLS06} which allows us to check for safety as well as liveness properties, thus removing an important limitation of the mentioned earlier work.

\section{Multithreaded BMC}\label{sec:multithread}
Our multithreaded environment is one where multiple solver threads $S= \{~ s_0,~s_1,~\dots,~s_n ~\}$ are run on a single shared memory system.
All the solver threads attempt to find a counterexample against property $\phi$ in model $M$, but they are not necessarily looking for counterexamples of the same length.
This means that in each solver thread $s_i$ the \emph{problem clause database} contains exactly the clauses in $\FM{sbnd(s_i)}$ for some bound $sbnd(s_i)$, the \emph{solver bound}.
Furthermore, let $minbnd(S) = min \{~ sbnd(s_i) ~|~ s_i \in S ~\}$ and $maxbnd(S) = max \{~ sbnd(s_i) ~|~ s_i \in S ~\}$ be the smallest respectively the largest solver bound amongst any of the solver threads in $S$.

Let $\LD{i}$ be the \emph{learned clause database} of solver thread $s_i$.
By definition each clause in the learned clause database is implied by the clauses in the problem clause database, so for each $C_j \in \LD{i}$ it holds that $\FM{sbnd(s_i)} \models C_j$.

The \emph{shared clause database} is a data structure accessible by each solver thread for the purpose of sharing learned clauses between solver threads.

\subsection{Approaches}
In our framework solver thread $s_i$ attempts to solve the formula $\FM{sbnd(s_i)}$.
Two solver threads $s_i, s_j \in S$ may have the same solver bound, i.e.{\ }it may hold that $sbnd(s_i)=sbnd(s_j)$, in which case both solver threads are solving the same formula.
A related approach in which no two threads are ever searching for a counterexample of the same length is presented in \cite{DBLP:conf/fmics/AbrahamSBFH06} for the checking of safety properties.
The restriction that no two threads must be solving the exact same formula may seem like it can only have positive effects, but this is not the case.
The reason is the lack of robustness of a SAT solving process.
Modern SAT solvers usually use some randomization, and due to this randomization the run time of a SAT solver may vary greatly for multiple runs of the same solver on the same formula when a different \emph{random seed} is used.
Recent work on distributed SAT solving \cite{JSAT:Hyvarinen,DBLP:conf/aisc/HyvarinenJN08} has confirmed that this can be exploited to achieve remarkable reductions in the expected run times by simply running the same randomized SAT solver on the same formula multiple times in parallel with different random seeds until one of them finishes.
By sharing clauses amongst these solver threads those results can be further improved.
The authors of \cite{JSAT:Hamadi} use a similar method for distributed SAT solving where they also consider using different search strategies in different threads (e.g.{\ }different solver parameter settings or even completely different SAT solvers).

A simple analogue to the described simple distribution methods for SAT that fits our framework is to make each solver thread independently act just like the conventional single-threaded approach \conv~that we described earlier.
We will call this approach \multiconv.

An approach similar to the one proposed in \cite{DBLP:conf/fmics/AbrahamSBFH06} in which each solver that has finished starts to search for a counterexample of the smallest length that no thread has started searching for (i.e.{\ }$maxbnd(S)+1$) we call \multibound.
In that approach the cores individually no longer follow the same scheme as \conv.

\subsection{Clause bound}\label{sec:ClauseBound}
For a clause $C_j$ let the \emph{clause bound} $cbnd(C_j)$ be a number such that $\FM{cbnd(C_j)} \models C_j$.
We use this clause bound for sharing learned clauses between solver threads.
The clause bound can be used to ensure that a solver thread $s_i$ only receives those shared clauses that are implied by the clauses in its problem clause database, as this holds at least for all clauses $C_j$ for which $cbnd(C_j) \leq sbnd(s_i)$.
To allow clause sharing whenever possible we would like $cbnd(C_j)$ to always be the \emph{minimal} bound at which $C_j$ is implied by the problem clauses, but this is hard to calculate and not required for correctness.
In fact, a safe approximation for the clause bound of any clause that is either in the problem clause database of solver thread $s_i$, or learned by that thread, would be $sbnd(s_i)$.

In our implementation we calculate a clause bound for each clause only once, after which it is stored with the clause.
With all clauses $C_j$ in the problem clause database we store $cbnd(C_j)~=~min \{~k~|~C_j \in \FM{k}~\}$, i.e.{\ }the first bound at which the clause appeared in the set of clauses.
Note that a learned clause is always derived from a number of other clauses.
For a learned clause $C_j$ derived from the set of clauses $\P$, we store $cbnd(C_j)=max \{~cbnd(C_k)~|~C_k \in \P~\}$, i.e.{\ }the maximum clause bound stored with any of the clauses in $\P$.
Finding the maximum clause bound of all clauses in the typically small set $\P$ takes only a negligible amount of time.

\subsection{Shared clause database organization}\label{sec:sharedclausedatabaseorganization}
The shared clause database is organized as a set of queues $\{~ Q_0,~Q_1,~\dots,~Q_{maxbnd(S)} ~\}$.
As the number of queues is dependent on $maxbnd(S)$ a new queue must be created whenever $maxbnd(S)$ increases.
This means that whenever a solver thread $s_i \in S$ starts to solve the problem for a bound that no other solver had reached up to that point it has to create a new queue in the shared clause database.

Each clause $C_j \in \LD{i}$ that solver thread $s_i$ wants to enter into the shared clause database should be pushed into queue $Q_{cbnd(C_j)}$.
Note that this is the queue corresponding to clause $C_j$'s \emph{clause bound}.
Each clause $C_j$ in queue $Q_k$ has a \emph{clause index} $q(Q_k, C_j)$.
The first clause to be pushed into an empty queue gets clause index $1$, and every clause pushed into a non-empty queue gets the number of its predecessor incremented by $1$.
Furthermore we define $p(Q_k,s_i)$ as the highest clause index amongst the clauses in $Q_k$ that solver thread $s_i$ knows about.
If solver $s_i$ has never read from nor written to queue $Q_k$ then $p(Q_k,s_i)=0$.

Each queue can be locked separately.
Furthermore there exists one readers-writer lock $\RWLock$ for the whole shared clause database.
A readers-writer lock can be acquired by multiple threads at the same time for reading or exclusively by one thread for writing.
If a thread wants to add a queue to the shared clause database it must acquire the lock $\RWLock$ for writing.
Threads that want to lock a separate queue for any type of access must first acquire lock $\RWLock$ for reading.
This mechanism is required because existing queues may be relocated in memory when a new queue is added to the database.

\begin{figure}[htp]
  \begin{example}Assume an environment in which two simultaneously working solver threads $S=\{~s_0,~s_1~\}$ exist, let $sbnd(s_0)=21$ and $sbnd(s_1)=22$.    
    A possible state of the shared clause database in this environment is the one depicted in Fig. \ref{fig:clausedb}.   
    The pointers $p(Q_{20},s_0)$ and $p(Q_{20},s_1)$ indicate that both solver threads have seen all clauses in queue $Q_{20}$.
    Solver thread $s_0$ has also seen all clauses from queue $Q_{21}$, but as its solver bound is smaller than $22$ it is not allowed to synchronize with queue $Q_{22}$ so it knows none of the clauses in there.    
    One may also observe that as solver thread $s_1$ has not seen the clauses $3-5$ in queue $Q_{21}$ they must have been put there by solver thread $s_0$.
  \end{example}
  \fig[width=.8\textwidth]{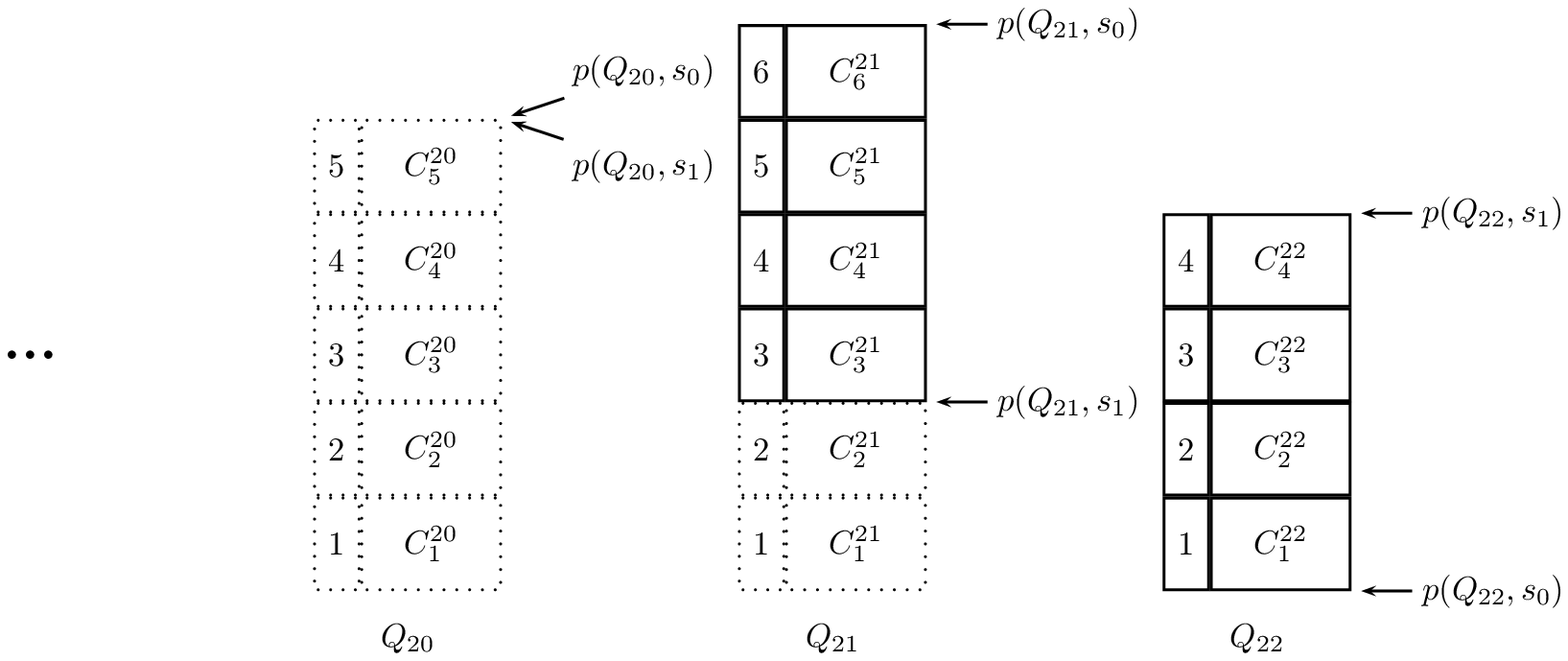}{Shared clause database example}
\end{figure}

\subsection{Synchronizing with the shared clause database}\label{sec:syncshared}
As explained in Subsection \ref{sec:ClauseBound}, all clauses $C_j$ for which $cbnd(C_j) \leq sbnd(s_i)$ are implied by the problem clauses in solver thread $s_i$, which means that $s_i$ can safely introduce all clauses from the queues $Q_k$ for $k \leq sbnd(s_i)$ to its shared clause database. As it does this it only has to read clauses it has not read before, so it can start reading from the clause with clause index $p(Q_k,s_i)+1$.

A clause $C_j$ can be removed from the queue $Q_k$ by the last solver thread $s_i \in S$ that reads it, i.e.{\ }when $s_i$ finds after reading that for all $s_m \in S$ it holds that $p(Q_k,s_m) \geq q(Q_k,C_j)$.
If a solver thread $s_i$ wishes to insert a set of clauses into queue $Q_k$ it must first lock that queue, then read all the clauses $C_j$ from it for which $q(Q_k,C_j) > p(Q_k,s_i)$.
Only after this it may write the new clauses to the queue and finally it may proceed to unlock it.
It is necessary that $s_i$ reads unread clauses from $Q_k$ before writing anything to it as otherwise the queue ends up in a state where clauses not known by $s_i$ precede clauses known by $s_i$.
In such a state we would no longer be able to use the clause index mechanism to identify which clauses in the queue the solver does not yet know.

Each solver thread $s_i \in S$ has a local clause stack $\LS{i} \subseteq \LD{i}$ that contains all clauses learned by $s_i$ that have not yet been placed in the shared clause database.
The clauses in stack $\LS{i}$ can be moved to the shared clause database at regular intervals.
As we have to read clauses from the database before writing to it, these points form the synchronization points of solver thread $s_i$ with the shared clause database.
The pseudocode for the synchronization procedure is stated in Algorithm \ref{syncalg}.
We chose to execute this synchronization at every \emph{restart} (see e.g.{\ }\cite{DBLP:conf/dac/MoskewiczMZZM01}), as restarts happen regularly but only after learning a substantial amount of new clauses, and because they are good points for introducing new learned clauses as all assignments of branching variables are undone.

\begin{algorithm}Synchronizing solver thread $s_i$ with the shared clause database.
  \normalfont{
    \begin{tabbing}
      1.~~ \= \LOCK~readers-writer lock $\RWLock$ for \textbf{reading} \\
      2. \> \FOR~\= all~$Q_k$~such that~$k \leq sbnd(s_i)$ \\
      3. \> \> \LOCK~queue $Q_k$ \\
      4. \> \> Read clauses $\{~ C_j ~|~ C_j \in Q_k,~q(Q_k,C_j) > p(Q_k,s_i) ~\}$ from the database \\
      5. \> \> Push clauses $\{~ C_j ~|~ C_j \in \LS{i},~cbnd(C_j) = k ~\}$ into $Q_k$ \\
      6. \> \> $newmin := min \{~ p(Q_k,s_m) ~|~ s_m \in S ~\}$ \\
      7. \> \> Remove all clauses $\{~ C_j ~|~ C_j \in Q_k,~q(Q_k,C_j) \leq newmin ~\}$ from the database \\
      8. \> \> \UNLOCK~queue $Q_k$ \\
      9. \> \ENDFOR \\
      10. \> \UNLOCK~readers-writer lock $\RWLock$ \\
      11. \> $\LS{i} := \emptyset$
    \end{tabbing}
  }
  \label{syncalg}
\end{algorithm}

As an optimization to this basic scheme our implementation pushes clauses $C_j$ for which $cbnd(C_j) < minbnd(S)$ into $Q_{minbnd(S)}$ instead of into $Q_{cbnd(C_j)}$.
This means that no clauses are pushed into queues corresponding to bounds that are no longer being solved by any solver thread. As a result the queues $Q_k$ for $k<minbnd(S)$ will eventually become empty after which they may be completely discarded.

\subsection{Benchmarks}\label{sec:multithreadbenchmarks}
We obtained the benchmark set used in \cite{DBLP:journals/lmcs/BiereHJLS06}, to which we will refer as LMCS06, and the benchmark suites L2S, TIP and Intel from the set of benchmarks used for the Hardware Model Checking Competition in 2007 (HWMCC07) \cite{HWMCC07}.
Each of the benchmarks represents a model $M$ and property $\phi$, which can serve as input to, for example, the model checker NuSMV \cite{DBLP:conf/cav/CimattiCGGPRST02}.

This model checker includes an implementation of the encoding presented in \cite{DBLP:journals/lmcs/BiereHJLS06}.
Unfortunately NuSMV is linked to an incremental SAT solver directly (e.g.{\ }MiniSAT) and thus the actual encoding of a benchmark into clauses that are fed to that solver does not become visible to its users.

We use a modified version of NuSMV version 2.4.3 which streams the sequence of formulas encoding a benchmark into a file rather than attempting to solve those formulas with its linked-in SAT solver.
For benchmarks from HWMCC07 for which it was known beforehand that the shortest existing counterexample was of length $k$, a formula sequence of length $k+11$ was generated, i.e.{\ }the largest formula represented in the file corresponds to the existence of a counterexample of length $k+10$.
For all other benchmarks the sequence was generated up to length $501$, i.e.{\ }the largest formula represented in the file corresponds to the existence of a counterexample of length $500$.
As no suitable file format existed for these incremental SAT problems we defined our own format, called \emph{\icnf}\footnote{For a detailed description, and tools for handling iCNF files, please check \url{http://www.tcs.hut.fi/~swiering/icnf/}}.

All of the obtained benchmarks were translated into a sequence of formulas as described.
\icnf~is \tarmo's input file format, so in the remainder of this paper whenever we speak of a benchmark we mean these translations.
We consider a benchmark solved when a formula in the sequence is found satisfiable, which corresponds to the existence of a counterexample, or when all formulas in the sequence are found unsatisfiable, which corresponds to the nonexistence of a counterexample of length at most 500.
We removed all benchmarks from our benchmark set that can be solved within 10 seconds by the single-threaded \conv~approach.
The resulting set contains $134$ benchmarks.

\subsection{Experiments}\label{sec:multithreadedexperiments}
In this subsection we present experimental results with different approaches to exploiting multi-core environments for BMC.
All results in this subsection were obtained using a single workstation from the set of 20 workstations found in our department's cluster.
Each workstation is equipped with two Intel Xeon 5130 (2 GHz) Dual Core processors and 16 GB of RAM.

Figures \ref{fig:cactus-1} and \ref{fig:cactus-2} are ``cactus plots'': such plots are traditionally used by the organizers of the SAT competitions \cite{satcompetitions} for comparing SAT solvers.
In a cactus plot, time is on the vertical axis and the number of instances solved is on the horizontal axis.
From Fig.~\ref{fig:cactus-1} one can, for example, see that for $97$ benchmarks in the set the run time of \conv~is under twenty minutes, and that for $105$ benchmarks the run time of \conv~is under one hour.

\begin{figure}[htp]
  \fig{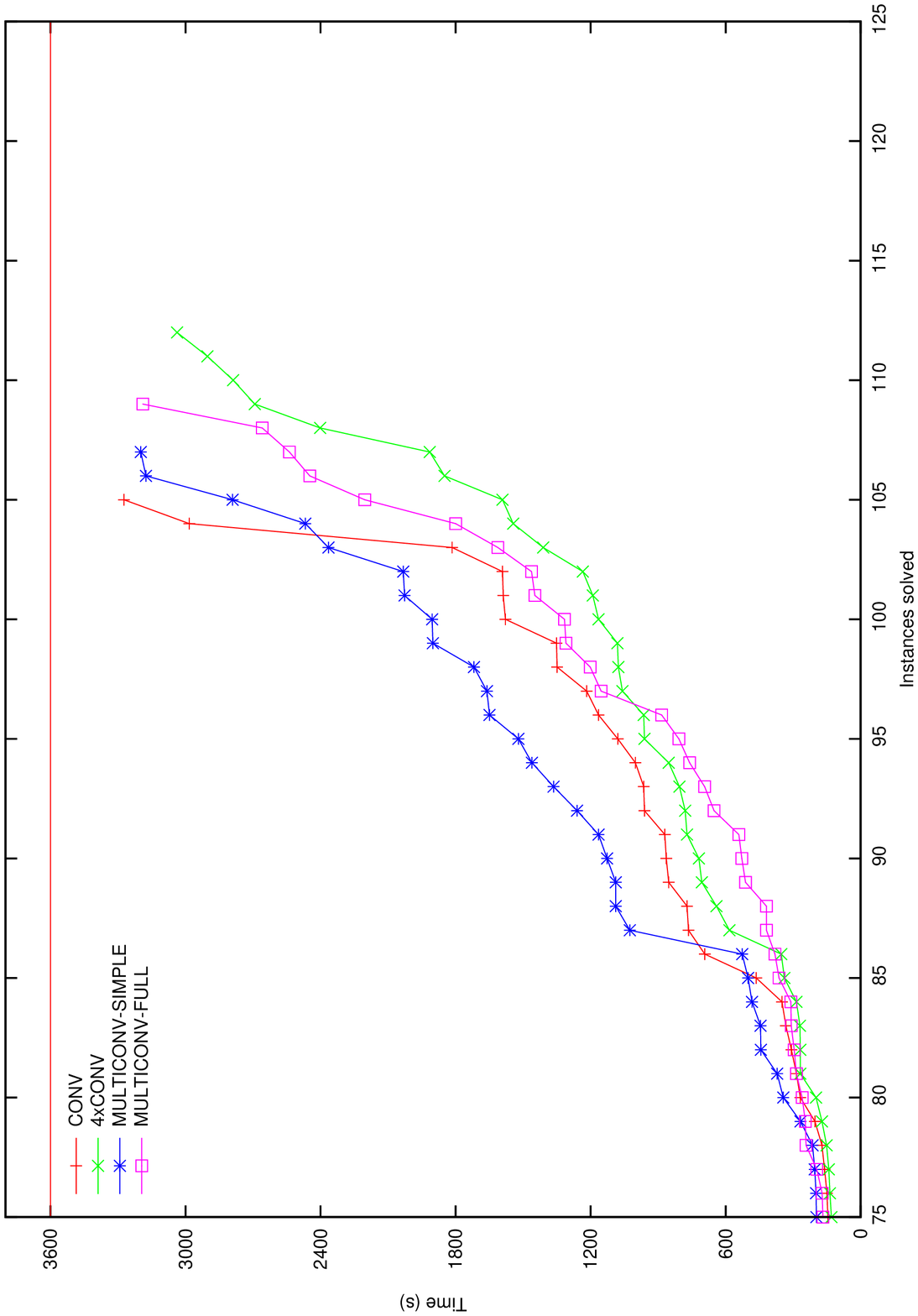}{Cactus plot showing the effects of multithreading.}
\end{figure}

\begin{figure}[htp]
  \fig{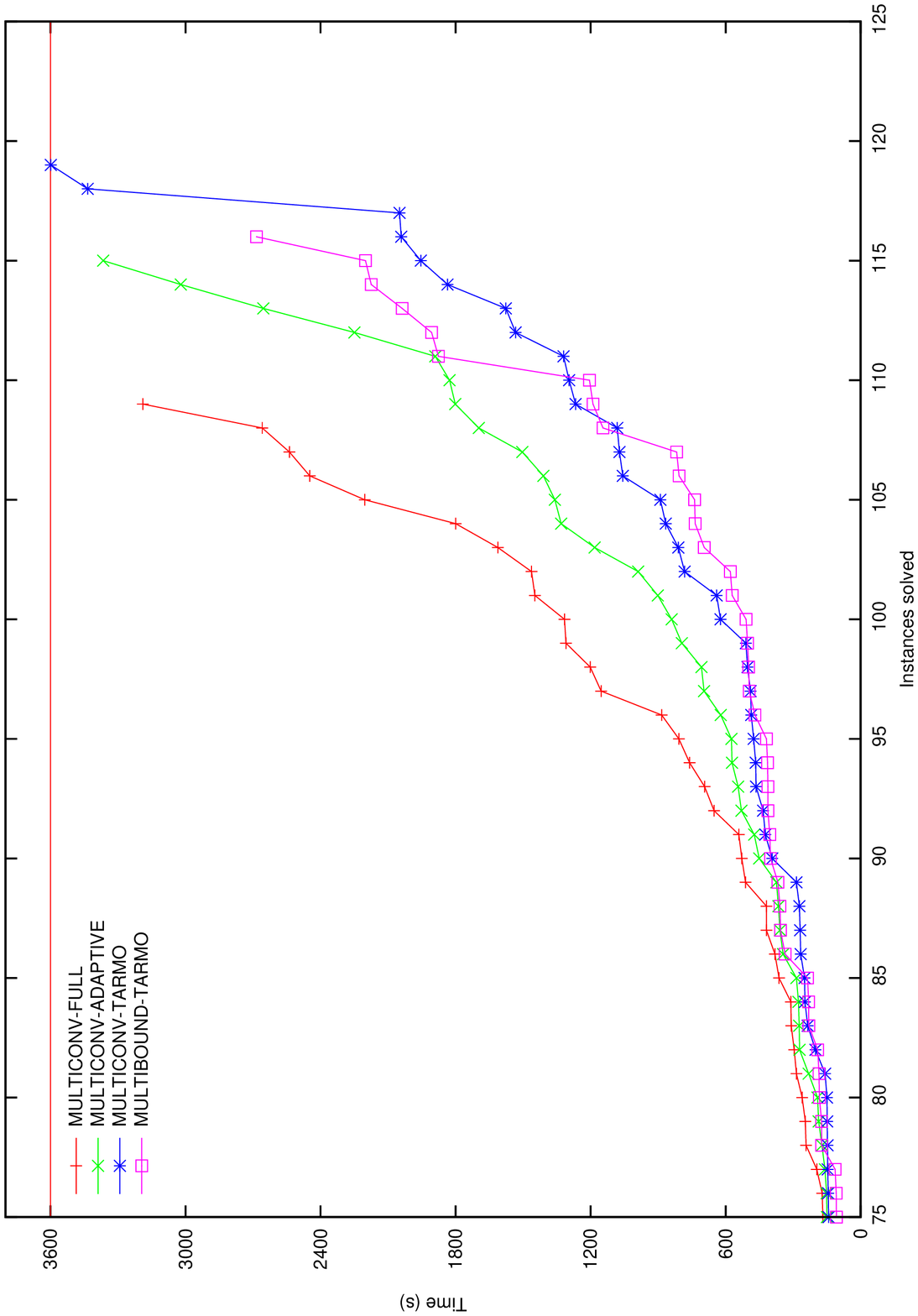}{Cactus plot showing the improved multithreaded variants.}
\end{figure}

\begin{figure}[htp]
  \fig{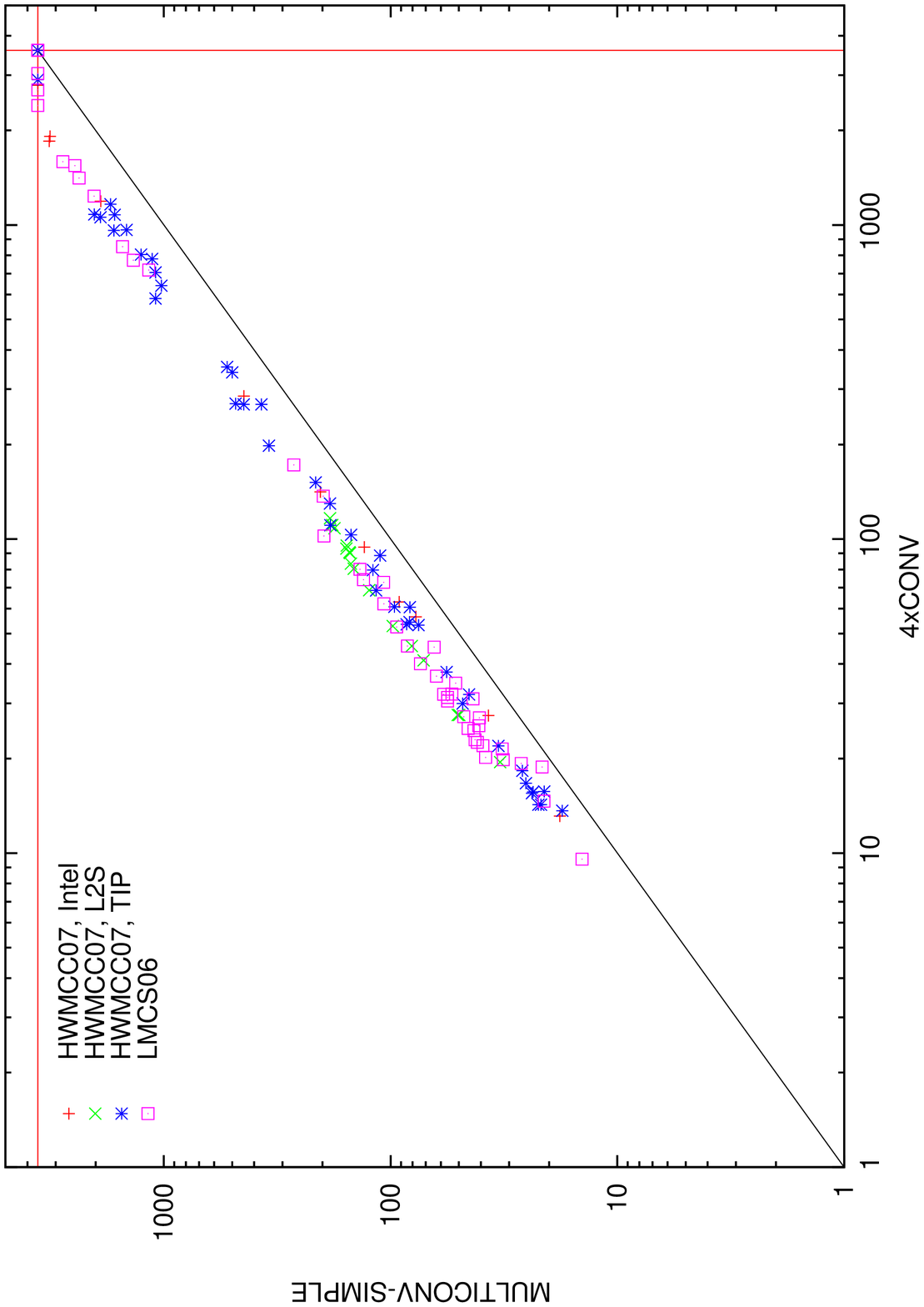}{Scatterplot illustrating the artificial variant \fourconv.}
\end{figure}

\begin{figure}[htp]
  \fig{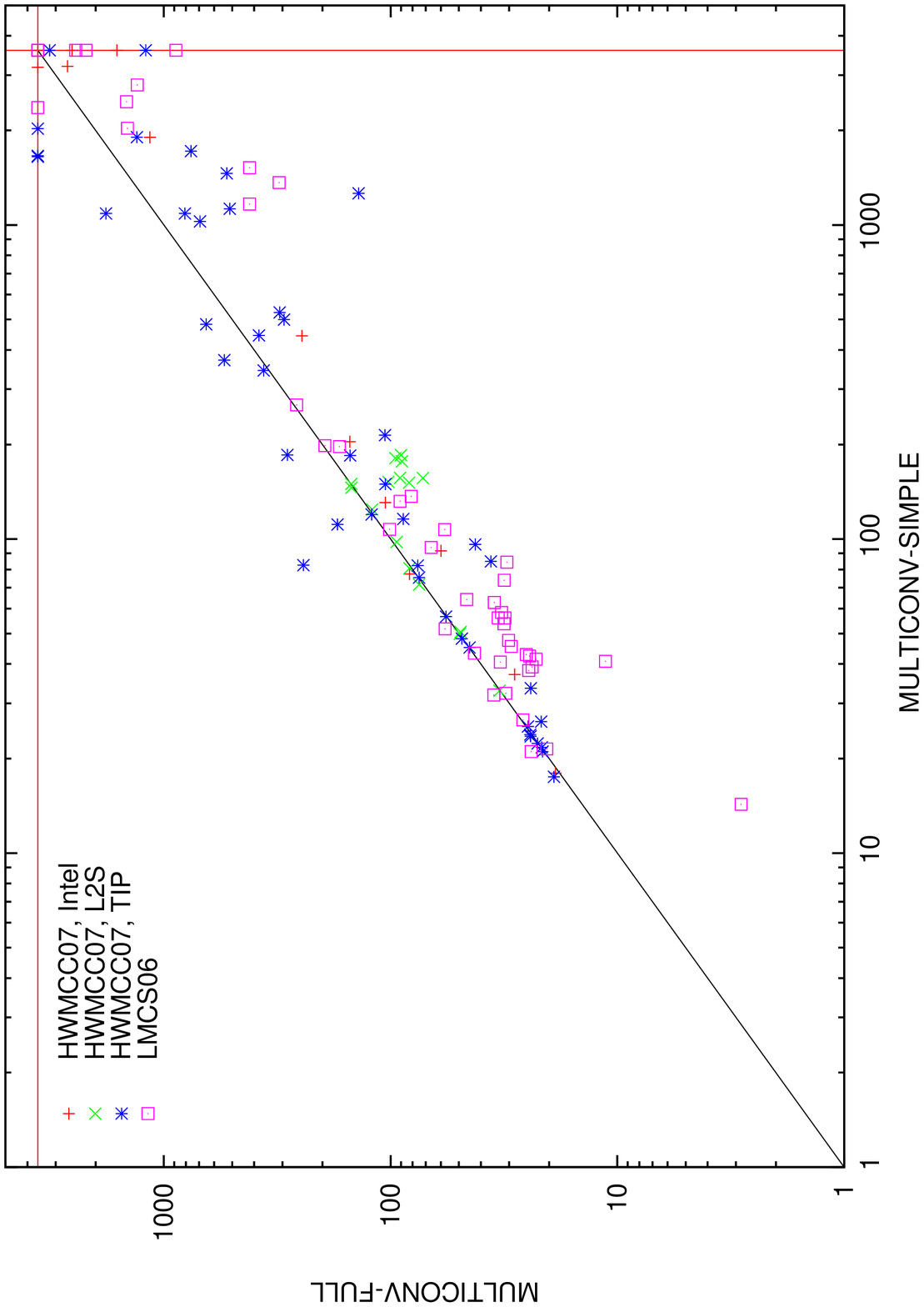}{Scatterplot illustrating the effect of clause sharing.}
\end{figure}

\begin{figure}[htp]
  \fig{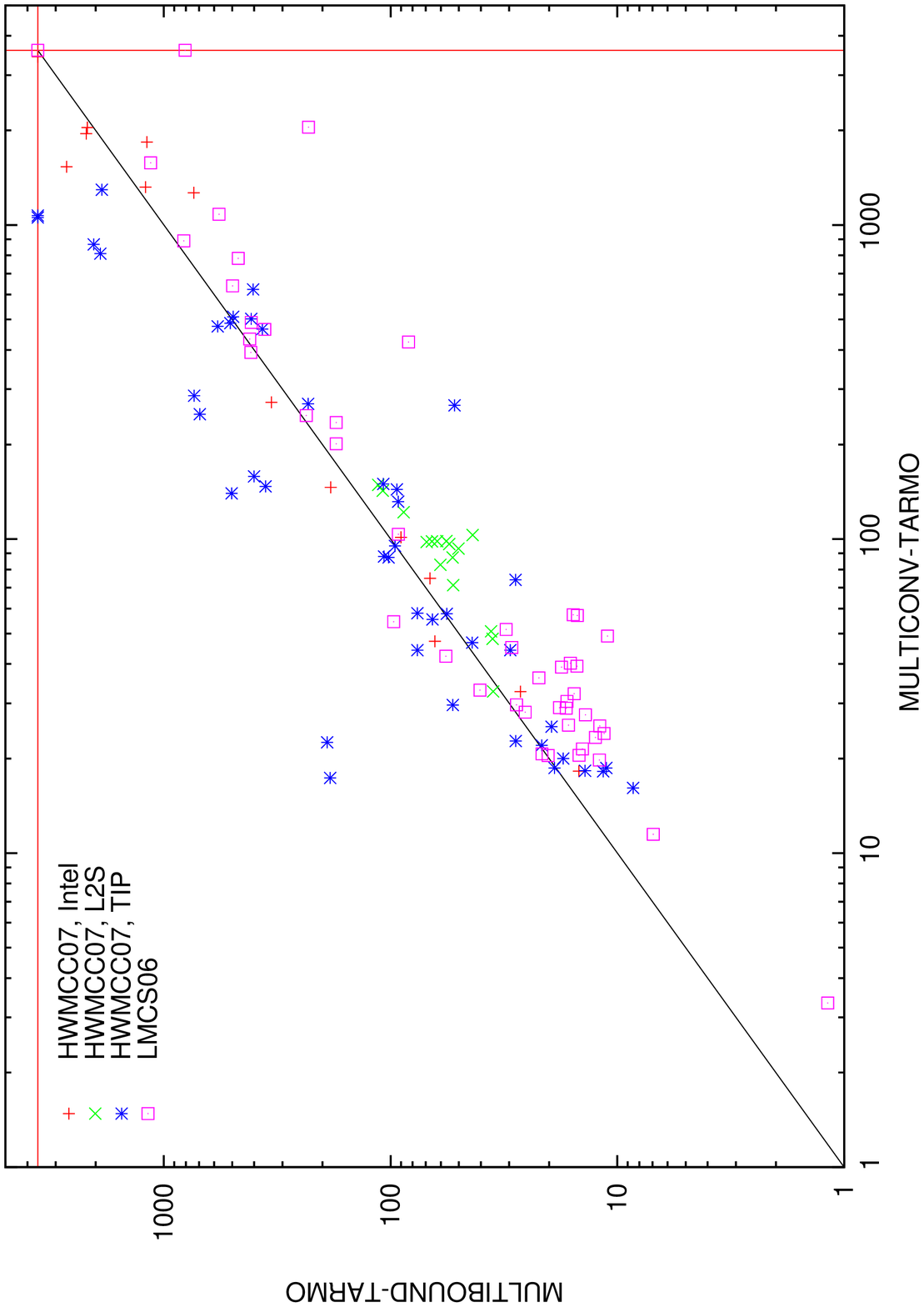}{Scatterplot comparing \multiconv~with \multibound.}
\end{figure}

The execution of the single-threaded \conv~obviously required the use of only a single core of one of our workstations, but, as will become clear later, it is important to note that care was taken to keep the other three available cores in that same workstation idle.
The results presented for \conv~are the run times of a single execution, but \conv~was executed in total four times for each benchmark.
\fourconv~is an artificial variant that reports the fastest of those four results for each benchmark.
This is meant to illustrate how the run time of a SAT solver varies per run due to the random choices it makes, and how this can be exploited to achieve reductions in the expected run time, as can be clearly seen from Fig.~\ref{fig:cactus-1}.

Unfortunately if we execute the four independent runs of \conv~in parallel on the same four core workstation the results are not as positive.
This is because the cores slow each other down as they share resources like the memory bus and parts of the cache.
The negative result can be clearly seen in the scatterplot presented in Fig.~\ref{fig:4xconv-vs-multiconv-simple} as well as in the cactus plot presented in Fig.~\ref{fig:cactus-1}.
From that cactus plot it can be seen how the result of this naive parallelization, which we will refer to as \multiconvsimple, is even slower than the single-threaded variant \conv~for many of the simpler benchmarks.
However, it does manage to solve a couple of benchmarks that \conv~could not solve within an hour.

Fortunately we can extend \multiconvsimple~with clause sharing to improve its performance.
\multiconvfullsharing~is a version which implements shared clause database synchronizations by every solver thread as described in Subsection \ref{sec:syncshared}.
Although one can see from the cactus plot presented in Fig.~\ref{fig:cactus-1} that the average performance improves after adding clause sharing, the scatterplot in Fig.~\ref{fig:simple-vs-full} shows that sharing clauses sometimes harms performance.
This was not unexpected as too many learned clauses are not beneficial to any SAT solver.
In fact, to reduce the negative effects of large learned clause databases SAT solvers occasionally delete learned clauses.

In distributed SAT solvers various ways of limiting the number of shared clauses can be found.
A common approach, found for example in \cite{JSAT:Hamadi}, is to share only clauses whose length is shorter than some constant.
This crude approach is justified by the observation that shorter clauses represent stronger constraints.
We have tried several such constants in our distributed BMC framework but we achieved better average results with variant \multiconvlesssharing~which uses an adaptive heuristic to limit clause sharing.
It shares only clauses whose length is smaller than or equal to the continuously recalculated average length of all clauses it ever learned.
The performance improvement can be clearly seen in Fig.~\ref{fig:cactus-2}.

In all of our \multiconv~variants presented so far the search space is pruned differently on each core only because of the effect of the randomization used by the SAT solvers.
To force a more diversified search we can use different search parameters in different threads.

One of MiniSAT's search parameters is the \emph{polarity mode} which can be either $negative$ or $positive$.
The default is $negative$, meaning that for every branching variable MiniSAT tries to assign the value $\false$~first.
In any case, MiniSAT selects the same value first consequently for each branching variable, which seems to be surprisingly effective \cite{SATAtDelft:hybrid}.
The default polarity mode $negative$ works best in practice for ``industrial'' SAT instances, which is solely caused by the way people tend to encode their problems.

We obtained the best results in our four-threaded environment with a variant we call \multiconvtarmo.
It is the same as \multiconvlesssharing~except that in one of the four solver threads we use the polarity mode $positive$.
This further diversifies the search, which causes a clear improvement of the performance as can be seen from Fig.~\ref{fig:cactus-2}.
Using polarity mode $positive$ in two of the four solver threads performed less well for our benchmarks.

We have also tested the \multibound~approach. 
Just as for \multiconv~we tested variants using full clause sharing, using our adaptive clause sharing heuristic, and with one solver using the opposite polarity mode setting.
In the cactus plot presented in Fig.~\ref{fig:cactus-2} only this last variant, called \multiboundtarmo, is plotted.
One can see that this version performs on average quite similarly to the equivalent \multiconv~variant.
Surprisingly enough the average performance of each \multibound~variant was similar to that of the equivalent \multiconv~variant.
This similar average performance is especially interesting since the performance for individual benchmarks is very different, as can be seen from the scatterplot presented in Fig.~\ref{fig:multiconv-vs-multibound}.
It thus seems that the \multiconv~and \multibound~approach are both useful, but complementary, approaches.

\section{BMC for workstation clusters}
Now that we have demonstrated the significant speed-ups that we can obtain using our multithreaded variants of \tarmo~we will discuss approaches which distribute runs of \tarmo~over several multithreaded workstations.
A distributed SAT solver for a similar environment is presented in \cite{JSAT:Schubert}.
The workstations in our department's computing cluster that were already mentioned in Subsection \ref{sec:multithreadbenchmarks} are all connected by 1 gigabit Ethernet connections through a cluster switch.

Our environment can be defined as a set $T= \{~ \dip,~S_0,~S_1,~\dots,~S_n ~\}$ in which $\dip$~refers to the single-threaded \emph{Database Interface Process} (DIP), and each $S_i$ is a \emph{worker}, which is simply a set of solver threads on a single multi-core workstation as defined in Section \ref{sec:multithread}.
Each multithreaded environment $S_i$ uses one of our multithreaded \tarmo~variants to find a counterexample against property $\phi$ in model $M$.

The DIP is a process which stores the \emph{global shared clause database}, and provides an interface to it for the solver threads.
It does not manipulate the database by itself.

For the remainder of this section let $Q^i_k$ refer to queue $Q_k$ in the local shared clause database of worker $S_i$, and $\GQ{k}$ refer to $Q_k$ in the global shared clause database stored in the DIP. 
Furthermore, let $\RWLock_i$ be the readers-writer lock for the local shared clause database of worker $S_i$.

\subsection{Global shared clause database organization}

The global shared clause database is a data structure which is almost identical to the shared clause database found in each worker process.
The difference is that it is accessed by the workers, rather than by their individual solver threads.
For each queue-worker pair $(\GQ{k},S_i)$ the clause database stores $p(\GQ{k},S_i)$ which is the highest clause index of the clauses in $\GQ{k}$ which worker $S_i$ knows about.

Only one worker can access the global shared clause database at the same time because the DIP is single-threaded.
This simplifies the design as well as preventing possible network congestion due to multiple workers accessing the database simultaneously.

\subsection{Global database synchronization}

Whenever a worker wishes to share clauses with other workers, one of its threads
performs a synchronization with the global shared clause database through the DIP.
This synchronizes the worker's local shared clause database with the global shared clause database.

Recall from Subsection \ref{sec:sharedclausedatabaseorganization} that we have for each thread $s_m \in S_i$ and queue $Q_k^i$ a clause index $p(Q_k^i,s_m)$.
The local database of each worker $S_i$ is extended with $p(Q_k^i,\dip)$ for each queue $Q_k^i$, where $p(Q_k^i,\dip)$ is defined as the highest clause index amongst all clauses in $Q_k^i$ that are known to the DIP.

The synchronization process begins with a worker $S_i$ sending a message to the DIP,
informing it that it is prepared for a synchronization.
The DIP gathers for all $\GQ{k}$ the clauses $\{~C_j ~|~ C_j \in \GQ{k},~q(\GQ{k},C_j) > p(\GQ{k},S_i) ~\}$ and places all of them in a buffer.
The whole buffer is then sent to worker $S_i$ at once.

When the worker has received the clause buffer from the DIP it starts a synchronization
procedure which is described in Algorithm \ref{global-syncalg}. As with local synchronizations, care
must be taken to ensure that writing new clauses to a queue always follows a lock and a read,
in order to prevent unknown clauses preceding known clauses in the queue.

\begin{algorithm}
Synchronizing worker $S_i$ with the global shared clause database.
\normalfont{
  \begin{tabbing}
     1.~~ \= Let $R$ be the set of clauses received from $\dip$ \\
     2. \> $B := \emptyset$ \\
     3. \> \LOCK~readers-writer lock $\RWLock_i$ for \textbf{reading} \\
     4. \> \FOR~\= all~$Q_k^i$~such that~$k \leq maxbnd(S_i)$ \\
     5. \> \> \LOCK~queue $Q_k^i$ \\
     6. \> \> Read clauses $\{~C_j ~|~ C_j \in Q_k^i,~q(Q_k^i,C_j) > p(Q_k^i,\dip) ~\}$ and append them to $B$ \\
     7. \> \> Push clauses $\{~C_j ~|~ C_j \in R,~cbnd(C_j) = k ~\}$ into $Q_k^i$ \\
     8. \> \> $newmin := min \left ( \{~ p(Q_k^i,s_m) ~|~ s_m \in S_i ~\} ~\cup~ \{~p(Q_k^i,\dip)~\} \right )$ \\
     9. \> \> Remove all clauses $\{~ C_j ~|~ C_j \in Q_k^i,~q(Q_k^i,C_j) \leq newmin ~\}$ \\
    10. \> \> \UNLOCK~queue $Q_k^i$ \\
    11. \> \ENDFOR \\
    12. \> \UNLOCK~readers-writer lock $\RWLock_i$ \\
    13. \> Send $B$ to $\dip$
  \end{tabbing}
}
\label{global-syncalg}
\end{algorithm}

Upon receiving the worker's learned clauses after the local synchronization has taken place, the DIP can write them to the global shared clause database. The process is completed and the DIP awaits another request.

\subsection{Experiments}
We have tried several approaches to distributing \tarmo~over more than one workstation.
Our best multithreaded variants turned out to be very robust. 
Simply running the same multithreaded variant multiple times with different seeds in parallel on several workstations and reporting the result when the first one finishes hardly decreases the expected run time.
From the experiments in Subsection \ref{sec:multithreadedexperiments} we concluded that our \multiconvtarmo~and \multiboundtarmo~variants both have good average performance but are complementary.
This observation inspired us to a simple distribution over two workstations where the two different approaches are each run on a single workstation.
In this way we obtain a result for each benchmark in exactly the amount of time it takes for the fastest of the two to finish.
We have named this variant \multiconvbound.
It was calculated from the earlier single workstation results rather than actually executed on two workstations in parallel. 
In this case this should, however, not make any difference to the result, as two workstations can function completely independently, at least assuming that they both already have the input file stored locally before starting the run.
From Fig.~\ref{fig:cactus-3} an improvement on the number of instances solved within an hour can be seen.
When one takes another look at Fig.~\ref{fig:multiconv-vs-multibound} in Section \ref{sec:multithreadedexperiments} one realizes that for many individual benchmarks the speed-up is significant as the achieved performance is the best of the two variants plotted there.

The cactus also shows the variant \distributed.
This is a truly distributed program that uses MPI version 2.0 \cite{grop99a} for communication between workstations.
To obtain each result for that variant we used three workstations in total: one running \multiconv, one running \multibound, and one running the DIP.
The single-threaded DIP was run on a single workstation in which the other three available processor cores were kept idle for the purpose of obtaining these results.
In a practical setting one will most likely not want to reserve an entire workstation for the single-threaded DIP, but as the DIP's computational load is not very high, relaxing that restriction should not cause a significant performance decrease.
It may even be a good choice in practice to run the DIP on the cluster's \emph{front-end}, which in a typical cluster setup is a single workstation through which all communication with machines outside the cluster takes place.

Note that in variant \distributed~we use the global shared clause database stored in the DIP to share clauses between a workstation running \multiconvtarmo~and a workstation running \multiboundtarmo. 
Our clause database design ensures that this does not cause any complications.
After testing several approaches we chose to have a worker initiate a synchronization with the global shared clause database whenever one of its solver threads increases its solver bound, i.e.{\ }every time a solver thread finds a formula unsatisfiable.
From Fig.~\ref{fig:cactus-3}~it can be seen that this simple global clause sharing setup improves the average performance.

This performance can probably be improved more by introducing a clever heuristic for limiting the number of clauses shared as we did for the multithreaded approaches. 
We chose not to further investigate such variants in this paper.
The performance increase obtained is mainly due to using two complementary multithreaded approaches.
As those are very robust approaches the performance of this distributed version of \tarmo~will not scale beyond two workstations.
One could try to define more multithreaded approaches with good average performance to obtain more complementary approaches that can be run in parallel but this is unlikely to scale much further.

This distributed framework with its generic shared clause database architecture will be very useful to our future work.
We plan to investigate approaches that use search space splitting amongst the workstations, in order to allow our system to scale to larger numbers of workstations.
A possible way of doing this would be to split the formulas using \emph{guiding paths} \cite{PSATO}.

\begin{figure}[htp]
  \fig{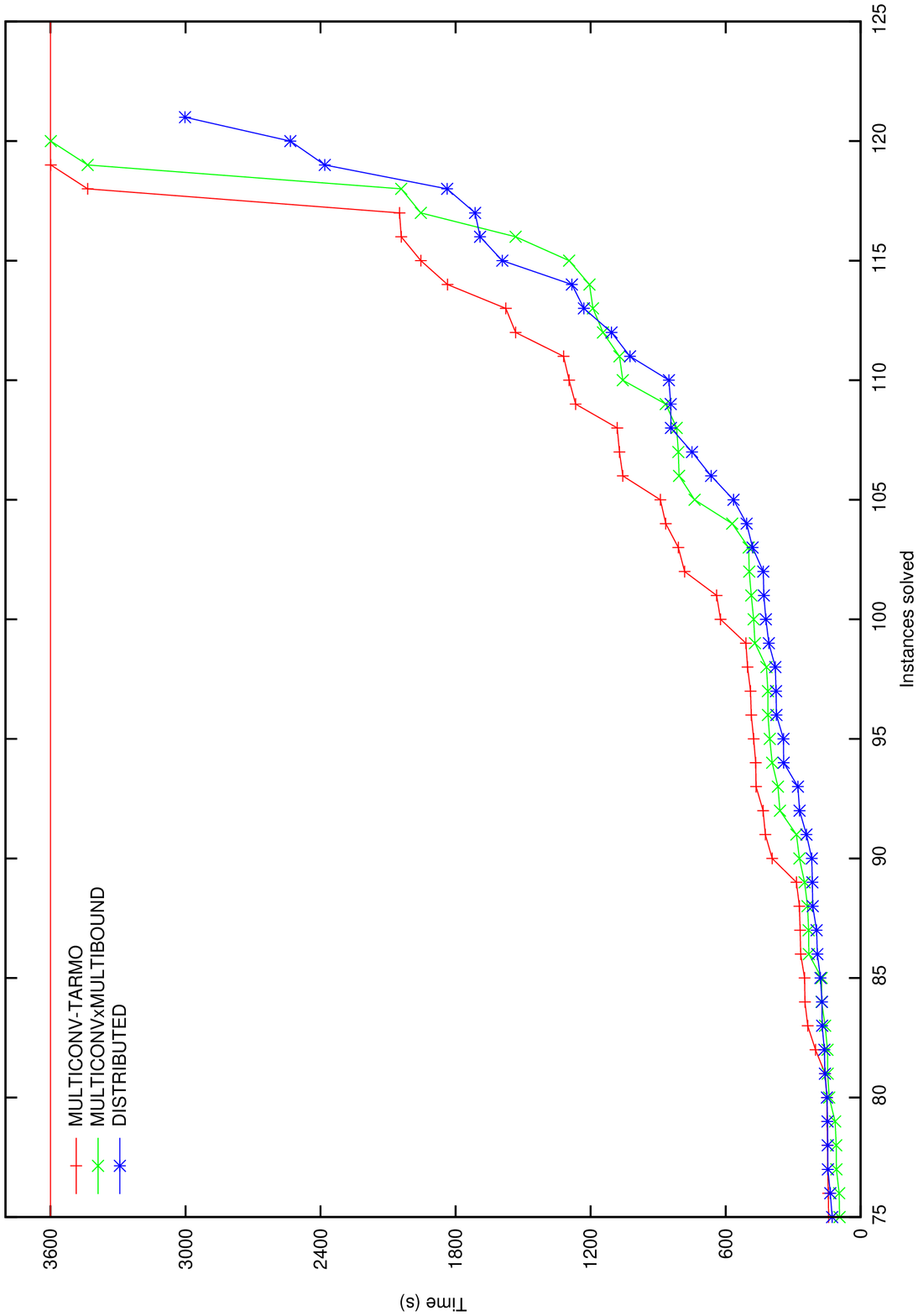}{Performance of the multiple workstation \tarmo~variants.}
\end{figure}

\section{Conclusion}
In this paper we have presented the \tarmo~framework for bounded model checking using multi-core workstations as well as clusters of them.
One novel feature of our framework for distributed BMC is that it allows using any encoding of BMC instances into incremental SAT. 
In our experiments we use the encoding presented in \cite{DBLP:journals/lmcs/BiereHJLS06}, which means that we are able to check safety as well as liveness properties with all variants of \tarmo~discussed in this paper.

An important contribution found in this work is our generic architecture for a shared clause database for multiple incremental SAT solver threads working on parts of the same incremental SAT encoding of a BMC instance.
Together with our definitions for \emph{clause bound} and \emph{solver bound}, it allows the sharing of clauses while requiring very little bookkeeping to make sure that solver threads only obtain those clauses that are are actually implied by their set of problem clauses.
It has been demonstrated how the architecture can be employed for solver threads operating in shared-memory environments as well as for solver threads that communicate through a network using MPI.

Our multi-core variants of \tarmo~obtained good speed-ups over the conventional single-threaded approach.
This is an important result as multi-core hardware is now widely available, and thus many BMC users can benefit from this.
Furthermore the two multi-core variants presented as \multiconvtarmo~and \multiboundtarmo~turned out to be complementary approaches which both have good average performance.

We exploited these complementary variants in a setting which uses multiple workstations.
We obtained a speed-up over the single workstation versions, but possibly more interestingly showed the feasibility of clause sharing between workstations using our shared clause database architecture.
This will be a very useful result for future distributed versions of \tarmo~or even other distributed BMC approaches.
To improve the rate at which the performance scales with the number of workstations used such future versions may, for example, split the search space into multiple disjoint parts.
Such techniques are easy to implement within our framework, as our shared clause database architecture allows clause sharing between any solver thread that is working on parts of the same incremental SAT problem, regardless of the solving strategy it uses.

Our \tarmo~implementation is available at: \url{http://www.tcs.hut.fi/~swiering/tarmo/}.

\vfill
\noindent \textbf{Acknowledgements} This work was financially supported by the Academy of Finland (projects 126860, 128050), Technology Industries of Finland Centennial Foundation, Jenny and Antti Wihuri Foundation, and the Helsinki Graduate School in Computer Science and Engineering (Hecse).

\bibliography{tarmo}
\end{document}